\documentclass[a4paper,fleqn,usenatbib]{mnras}
\usepackage{amsmath,amstext}
\usepackage[T1]{fontenc}
\usepackage{ae,aecompl}
\usepackage{graphicx}	
\usepackage{amssymb}	
\usepackage{booktabs,caption}

\newcommand{\fesc}{f_{\text{esc}}}
\newcommand{\Tvir}{T_{\text{vir}}}
\newcommand{\Rf}{R_{\text{mfp}}}
\newcommand{\Zion}{\zeta_{\text{ion}}}
\newcommand{\ODGcentieme}{\,\times 10^{\,\text{-}2}}

\usepackage{float}

\title[EoR parameters reconstruction]{Improved supervised learning methods for EoR parameters reconstruction}

\author[Doussot et al.]{Aristide Doussot$^{1}$, Evan Eames$^{1}$ and Benoit Semelin$^{1}$
\\
$^{1}$Sorbonne Universit\'e, Observatoire de Paris, PSL research university, CNRS, LERMA, F-75014 Paris, France\\}

\begin{document}

\maketitle
\begin{abstract}

	Within the next few years, the Square Kilometer Array (SKA) or one of its pathfinders will hopefully provide a detection of the 21-cm signal fluctuations from the Epoch of Reionization (EoR). Then, the main goal will be to accurately constrain the underlying astrophysical parameters. Currently, this is mainly done with Bayesian inference using Markov Chain Monte Carlo sampling. Recently, studies using neural networks trained to performed inverse modelling have shown interesting results. We build on these by improving the accuracy of the predictions using neural network and exploring other supervised learning methods: the kernel and ridge regressions. Based on a large training set of 21-cm power spectra, we compare the performances of these supervised learning methods. When using an un-noised signal as input, we improve on previous neural network accuracy by one order of magnitude and, using local ridge kernel regression, we gain another factor of a few. We then reach a rms prediction error of a few percents of the 1-sigma confidence level due to SKA thermal noise (as estimated with Bayesian inference). This last performance level requires optimizing the hyper-parameters of the method: how to do that perfectly in the case of an unknown signal remains an open question. For an input signal altered by a SKA-type thermal noise, our neural network recovers the astrophysical parameter values with an error within half of the 1$\sigma$ confidence level due to the SKA thermal noise. This accuracy improves to 10$\%$ of the 1$\sigma$ level when using the local ridge kernel regression (with optimized hyper-parameters). We are thus reaching a performance level where supervised learning methods are a viable alternative to determine the best-fit parameters values.

\end{abstract}

\section{Introduction}

It has been recognized for more than 20 years that the neutral hydrogen in the Inter-Galactic Medium (IGM) before and during the process of reionization of the universe must have emitted radiations at $21$ cm that, redshifted at meter wavelengths by cosmic expansion, should be observable nowadays with adequate radiotelescopes \citep{Madau97, Furlanetto2006, Pritchard2012, Mellema13, Koopmans15}. The main difficulty in detecting this signal is to separate it from various types of foreground emissions \citep[galactic synchrotron, extragalactic point sources, etc., see e.g.][]{Dimatteo04,Jelic08}. 
Single dipole instruments can measure the intensity of the signal as a function of frequency integrated on the sky (global signal). Such observations have the advantage that the signal-to-noise ratio does not depend on the collecting area, but the drawback that the limited amount of collected information gives us less leverage to separate the signal from the foregrounds and encodes less knowledge about the underlying astrophysical processes. A tentative first detection of the global signal has been reported by \citet{Bowman18} with the EDGES instrument. 
It is likely that the cosmic origin of the detected feature can only be ascertained with future interferometric observations that would quantify the angular fluctuations in the detected feature (in the form of a power  spectrum).

A number of instruments have been attempting to measure the power spectrum of the signal, although mostly at higher frequency (and thus lower redshift) than the EDGES detection. Only upper limits have been established so far, at various wavenumbers and redshifts \citep{Paciga13, Beardsley16, Patil17, Ali18}. With this type of observations, unambiguous separation of signal and foregrounds should be possible but the calibration of the instrument is a much more difficult task. Also, the higher information content of the measured quantity comes with the requirement of a large collecting area to improve the signal-to-noise ratio. As the methods to perform the calibration improve, we may see a first interferometric detection in the next few years. Then, next generation instruments such as SKA and HERA should be able to measure the power spectrum much more accurately, detect it at lower frequencies (higher redshifts) where the foregrounds are stronger, and even, in the case of SKA, image the signal in three dimensions.

There is obviously a trade off between the amount of information in a type of observation (global signal, power spectrum, imaging) and the collecting area and integration time required to perform it with a good enough signal-to-noise ratio. But in all cases, transforming this information into astrophysical or cosmological knowledge is not a straightforward process. Indeed, the local intensity of the signal depends, in some cases non-linearly, on the hydrogen density, ionization fraction, velocity, kinetic temperature and on the local Lyman-$\alpha$ radiation field \citep[see][]{Furlanetto2006}. These quantities are in turn correlated in a non-trivial manner through the process of structure formation. Thus the first crucial step in interpreting the signal is to build a model that can compute the signal from such basic processes as growth of density fluctuations, formation of sources of radiations (stars and AGN), and radiative feedback of the sources on their direct environment and on the IGM. These models can be analytical \citep[e.g.][]{Barkana05,Pritchard07}, semi-numerical \citep[e.g.][]{Thomas09,Santos10,Mesinger11,Fialkov14,Ghara15}, or the result of radiative transfer cosmological simulations \citep[e.g.][and subsequent works]{Gnedin04,Mellema06,Valdes06,McQuinn07b,Baek09}. In all cases the models will requires the use of astrophysical parameters to describe processes either not implemented {\sl ab initio} and\slash or below the resolution of the computation. A simple example is the efficiency of star formation, that would require a mass resolution below 1 solar mass and an extremely short time step to be modelled self-consistently.

 The second important step in extracting astrophysical knowledge from the observation, is to use reliable statistical methods to put constraints to the models astrophysical parameters. A number of such methods exists. \citet{Pober14} used the Fisher information matrix to derive confidence intervals for the parameter values, \citet{Greig15,Greig17,Greig18} use Bayesian inference enacted by Markov Chain Monte Carlo (MCMC) with the semi-numerical code 21cmFast. As even using 21cmFast for MCMC Bayesian inference is computationally expensive, \citet{Kern17}, \citet{Schmit18} and \citet{Jennings19} build an emulator of the code, using Gaussian Processes and Neural Networks. Another approach to parameter estimation is to train supervised learning algorithms to perform inverse modelling, taking some representation of the observed signal as an input and directly predicting the parameter values. As in the case of building an emulator, the training is specific to the chosen model, and typically requires a smaller number of modelling runs as MCMC inference does. For now, neural networks trained for inverse modelling have been implemented using the power spectrum \citep{Shimabukuro2017} or the full tomographic data \citep{Gillet18} as input, to predict best-fit parameters values. Predicting confidence levels could be done in various ways, for example using Bayesian neural networks. It is not obvious at this stage, however, that the predicted confidence levels would have the exact same meaning as in classical Bayesian inference.
 
 When predicting best-fit parameters using neural networks (or other supervised learning algorithms) trained to perform inverse modelling, an error exists, due to the imperfect training of the network. This training error can be exactly computed if the test input signal was produced by the model itself (then we know the corresponding {\sl true} parameters). Note that finding parameter values that do not perfectly match a test signal {\sl not} produced with the model is an issue not specific to supervised learning methods: maximum-likelihood parameters with a low likelihood value indicate an imperfect model. In any case,
 for supervised learning methods to be actually usable, we need to ensure that the training error is much smaller than the typical 1-sigma confidence due to the thermal noise in the target observation, as estimated by Bayesian inference. This should of course be true if an un-noise signal is fed to the network, but also if a noised signal is considered. Such was not really the case in \citet{Shimabukuro2017} where the error is of the same order as the thermal noise, or in \citet{Gillet18} where only an un-noised signal is considered. Thus we need to improve the performance of supervised learning method implementations, either by improving the implemented methods, or exploring new ones.

In this work we explore both of these possibilities. First, we improve substantially on the performances reached in \citet{Shimabukuro2017} using neural networks, by using a larger learning set, and optimizing several steps in the process. Then we explore another supervised learning method. Neural network have encountered great success when dealing with image classification. In this situation, the dimension of the signal space is huge, typically of the order of $10^6$, the number of pixels in the image. In our case, the signal is the value of the power spectrum at various wavenumbers and redshifts. The dimension of the signal space is much lower, typically of the order of $10^2$. In such comparatively low dimensions, advanced versions of the classical linear regression are known to perform well. Indeed, the linear regression, using the knowledge from a set of samples to approximate a model with a linear relation, can be classified as supervised learning.
As $10^2$ dimensions is still very large to apply the classical linear regression, kernel regression and ridge regressions have been developed \citep{Hastie2001}. In this work we combined these improved regression methods and push them as far as we can is term of performance to compare them with the neural network approach.

The layout of this article is as follows. In section 2 we present the case to which we apply supervised learning: the input signal and thermal noise, the parameters to be predicted and the model that relates them in the case of forward modelling. In section 3 we detail the different supervised learning methods studied in this work. In section 4 we study the accuracy the these methods in term of the error on the reconstructed parameter values. In section 5 we present our conclusions.

\section{Framework}

\subsection{The model: 21cmFast}

	The learning process of any supervised learning method requires a training set consisting of a sufficient number of labeled cases (where both inputs and outputs are known). Generating such a number of cases is currently  beyond the reach of full-numerical simulations designed to predict the 21-cm signal. Consequently, we have selected the semi-numerical code 21cmFast \citep{Mesinger2011} which is fast enough to provide the required number of cases in a reasonable amount of time. Let us briefly review some salient features of 21cmFast numerical methods.
	
	The main feature is that 21cmFast does not include full radiative transfer, thus saving a lot of computation time. Instead, the ionization process is based on the "excursion-set" approach \citep{Furlanetto2004, Mesinger2007}. The basic principle is that if the number of ionizing photon produced in a region is larger than the number of neutral hydrogen atoms in the same region, the region is considered ionized (in practice, only the region center cell is tagged as ionized, as regions centered on all cells will be considered). The photon production rate is assumed to be proportional to the collapse fraction (fraction of baryons in a collapsed object). At each location, the collapsed fraction smoothed on scale $R$, $f_{\text{coll}}(\mathbf{x},z,R)$, is compared to an efficiency parameter $\Zion$. The comparison is performed for decreasing $R$ values, from a large scale $\Rf$ to the cell size $R_{\text{cell}}$. If $f_{\text{coll}}(\mathbf{x},z,R)>\Zion^{-1}$ then the center cell of the region is flagged as ionized. Finally, at $R_{\text{cell}}$, the ionizing fraction of the remaining cells that are not fully ionized is set to be $\Zion f_{\text{coll}}(\mathbf{x},z,R_{\text{cell}})$. See \cite{Mesinger2011} for further details.
	
	Another cost-saving strategy implemented in 21cmFast is to ignore baryonic dynamics, and use simplified dark matter dynamics. The dark matter density field is linearly extrapolated from the primordial field using the standard Zel'Dovich approximation \citep{Zeldovich1970}. Baryons are simply assumed to track the dark matter exactly. See \citet{Mesinger2007} for further details. X-ray heating and Lyman-$\alpha$ contributions to the spin temperature of hydrogen are implemented in 21cmFast, again using cost-saving strategies. However we deactivated these processes in our study, setting ourselves in the high spin temperature limit. 

\subsection{Selected EoR observables and model parameters}
In our approach to supervised learning where our goal is to put constraints on model parameters using observables, the observables are the inputs of the method and parameters values are the outputs. Let us specify which inputs and outputs have been used in this work.
\subsubsection{EoR observable}
	In our study, we chose to focus on the power spectrum of the intergalactic 21-cm signal, assuming that the non-gaussianities \citep{Shaw2019} of the signal are not necessary to accurately reconstruct the parameters. More precisely we chose to consider the values of the power spectrum at 12 different wavenumbers $k$, logarithmically sampled from 4.42$\times 10^{-2}$ cMpc$^{-1}$ to 3.20 cMpc$^{-1}$, for integer values of the redshift $z$ from $5$ to $15$. Then, the signal that is used as an input lives in a space of dimension $120$. This choice allows us to work in relatively low dimension unlike, for example, \citet{Gillet2018} who deal with the full information from the lightcone using convolutional neural networks. While neural networks have shown their ability to deal with high dimensional signals (dimension $10^6$) when analyzing images for example, other supervised learning method, such as the different flavors of linear regression presented here, are well suited to lower dimensionality.

\subsubsection{Choice of EoR parameters and sampling}

	Concerning the EoR parameters that we want to reconstruct, we have chosen three parameters that have often been considered in other works \citep{Greig2015a, Greig2017a, Greig2018a, Schmit2018, Eames2018}:

\begin{itemize}
	\item $\Zion$ accounts for the ionizing efficiency of high-z galaxies and can be expressed as:
	\begin{equation}
		\Zion = 30\left(\frac{\fesc}{0.3}\right)\left(\frac{f_{*}}{0.05}\right)\left(\frac{N_{\gamma}}{4000}\right)\left(\frac{2}{1+n_{rec}}\right)
	\end{equation}
	with $\fesc$ the ionizing photon escape fraction, $f_{*}$ the fraction of galactic gas in stars, $N_{\gamma}$ the number of ionizing photons produced per baryon in stars, and $n_{rec}$ the typical number of times a hydrogen atom recombines during the EoR
	\item $\Rf$ is the mean free path of ionizing photons within the ionized regions, regulated by the existence of unresolved Damped Lyman-$\alpha$ systems.
	\item $\Tvir$ is a mass threshold above which halos are allowed to form stars and begin ionizing their surroundings.
\end{itemize}

	Detailed definitions of these parameters are given in \citet{Greig2015a}. 
	
	We based our study on a learning set of 2400 labeled cases, generated for our previous study in \citet{Eames2018}, corresponding to the nodes of a logarithmic 20$\times$6$\times$20 grid in the parameter space $(\Zion ; \Rf ; \Tvir )$ with the following boundaries :
\begin{itemize}
	 \item[$-$]$\Zion\in [20, 200]$\\
	 \item[$-$]$\Rf\in [5\text{ cMpc}, 35\text{ cMpc}]$\\
	 \item[$-$]$\Tvir\in [8.0\times 10^{3}\text{K}, 10^{5}\text{K}]$\\
\end{itemize}

	  Let us emphasize that this sampling method by no means ensures a maximization of the information. Methods that optimize the sampling (for a fixed number of cases and fixed explored volume in parameter space) to maximize the information are presented in \citet{Eames2018} and appear to lead to a better training of, at least, neural network methods. Further details on the setup of the 21cmFast runs performed for each triplet of parameter values can be found in \citet{Eames2018}.
	
\subsection{SKA noise modelling}

	For supervised learning methods designed to constrain model parameters to be of any use, they have to be able to handle a signal affected by the observational noise. We will concentrate on the thermal noise from the SKA, neglecting other possible sources such as imperfect foreground removal, residual calibration errors, or even sample variance. To model the expected thermal noise we consider the SKA specifications as detailed in \citet{Dewdney2013}. Following \citet{McQuinn2005}, we write the detector noise covariance matrix as:
	
\begin{equation}
	C(\mathbf{k}_i, \mathbf{k}_j)= \frac{1}{Bt_{\mathbf{k}_i}}\left( \frac{\lambda^{2}BT_{\mathrm{sys}}}{A_{e}}\right)^{2} \delta_{ij}
\label{Eq:DetectorNoise}
\end{equation}
where $B$ is the bandwidth, $t_{\mathbf{k}_i}$ is the effective observing time of the instrument in the grided visibility cell corresponding to wavenumber $\mathbf{k}_i$, $\lambda$  is the observed wavelength, $T_{\mathrm{sys}}$ is the total system temperature and $A_{e}$ is the effective area of the station. For the system temperature we have used $T_{\mathrm{sys}}= 100 + 300 \left( \frac{\nu}{150 \mathrm{MHz}} \right)^{-2.55}$K \citep{Mellema2013}. Lacking data from a definitive design of the future SKA-Low antennas, we have used an effective area for a station composed of 256 antennas of: $A_e=256 \times \mathrm{min}(2.56,{\lambda^2 \over 3}) \,\,\mathrm{m}^2$. We have assumed a bandwidth $B=10$ MHz, and a station diameter of $35$ m determining the field of view. Finally we have computed the $t_{\mathbf{k}_i}$ by integrating in visibility space the trajectories of the baselines from the SKA specifications. We considered 8h runs for a total integration time of 1000h, and a target field with declination $-30 \deg$ (close to the zenith for SKA-Low).

From the detector noise covariance matrix we can compute the 1$\sigma$ uncertainty on the power spectrum due to thermal noise as:
\begin{equation}
		\delta P^{21}_{\Delta T}\left( k\right) =\left[\sum\limits_{\left|\mathbf{k}\right|=k}\left(\frac{1}{	{A_e x^2 y \over \lambda^2 B^2} C(\mathbf{k},\mathbf{k})}\right)^{2}\right]^{-\frac{1}{2}}
\label{Eq:AverageThermalNoise}
\end{equation}
where the sum extends over Fourier-space cells in the spherical shell with radius $k$ (and also thickness $\Delta k=k$ in our case, which is a usual but determining choice), $x$ is the comoving distance to the observed redshift and $y$ the depth of the field (a distance) as determined by the bandwidth and the cosmology. The resulting level of noise is very similar to that in \citet{Koopmans15}.

	See \citet{McQuinn2005} for further details on establishing the formulas. Once $\delta P^{21}_{\Delta T}\left( k\right)$ is computed for our binned wavenumbers, we simply add a realization of this noise to the signal to get a noised power spectrum.

\section{Supervised learning methods}\label{Sec:Methods}

	A well established way of predicting underlying astrophysical parameters using observables is Bayesian inference associated to Markov Chain Monte Carlo sampling. However, it often requires numerous instances of forward modelling to predict one observable, like in 21CMMC \citep{Greig2015a,Greig2017a, Greig2018a, Park2018} where the forward modelling is performed using 21cmFast \citep{Mesinger2011}. This inherently comes with a high computational cost, even if some attempts on designing a fast 21-cm power spectrum emulator using gaussian processes \citep{Kern2017, Jennings2018} or support vector machine \citep{Jennings2018} to replace 21cmFast have significantly accelerated the process. 
	
	With supervised learning methods trained to perform inverse modelling and predict parameter values, a typically smaller number of forward modelling instances is needed to build the learning set, decreasing the required computational time compared to 21CMMC. We chose to focus on neural networks, as it appears to be the fastest method in term of computational time, and on ridge and kernel regressions which have not been explored for this purpose before.
	
\subsection{Common features}
\subsubsection{Learning set and test set}

	Although different, the two classes of supervised learning methods studied in this work share common features. In essence, the supervised learning material consists of a set of labeled cases, from which the algorithm can interpolate to successfully make predictions for cases not in the set. This set of labeled cases is usually called the learning set in the neural network field. To quantify the prediction quality of a method, a second set of labeled cases, distinct of the first one, is used. In the neural network field, this sample is often referred as the test set. Performing the evaluation on the test set avoids being impacted by the well-know issue of over-fitting on the learning set.
	
	In our study the learning set is either made of $2400$ signals for the cases without instrumental noise added \citep[that were already described in][]{Eames2018} or of $20$ noised realizations of each signals, which means $48 000$ noised signals, when instrumental noise is taken into account. The test set is composed of $512$ signals generated starting from random values of the three astrophysical parameters taken within the bounds of the grid-based learning set. When noise is included, we generate $10$ realizations of each signal which leads to a test set composed of $5120$ noised signals.
	
\subsubsection{Limitations to absolute performance evaluation}\label{SubSec:Evaluation_Issue}

	Any supervised learning algorithm includes, in various forms, adjustable quantities, often called weights, that encode the computation of the outputs. The learning process therefore consists on adjusting the weights to accurately recover the known (labeled) outputs of the learning set, based on its inputs, by minimizing a given error function. The function to minimize usually depends on various adjustable hyper-parameters like a learning rate $\eta$ or the weight decay rate $\lambda$. Changing the values of these hyper-parameters thus leads to different error functions, different minimization results and therefore different predictions. Optimizing the values of the hyper-parameters is of paramount importance to obtain the best possible predictions. However it is almost impossible to make sure that a set of hyper-parameters values is a global optimum, especially with neural networks where there is an infinite number of possible architectures. The comparison between methods can only be done with parameter values that are, at best, local optima in the hyper-parameter space.
	
	Also, our learning and test sets are generated with the same semi-numerical model: 21cmFast \citep{Mesinger2011}. Any conclusion that we reach concerning the accuracy
	of parameter reconstruction using different methods will only hold when applied to a real observed signal if the model is able to reproduce the observed signal. More quantitatively, in the 120-dimensional signal space, the signals produced by our 3-parameters model occupy a 3 dimensional manifold. The observed signal will not lie in this manifold unless the model is perfect. Even when noise is included, the performances of our parameter reconstructions methods are only evaluated close to this manifolds (at distances typically corresponding to a 1-sigma thermal noise). If the observed signal is at a distance equivalent to many sigmas, our conclusions cannot apply.

\subsection{Preparing the data}
\subsubsection{Labelling a noised signal: theoretical issue}\label{SubSec:PerpTh}

\begin{figure}
  \centering
  \includegraphics[width=.45\columnwidth]{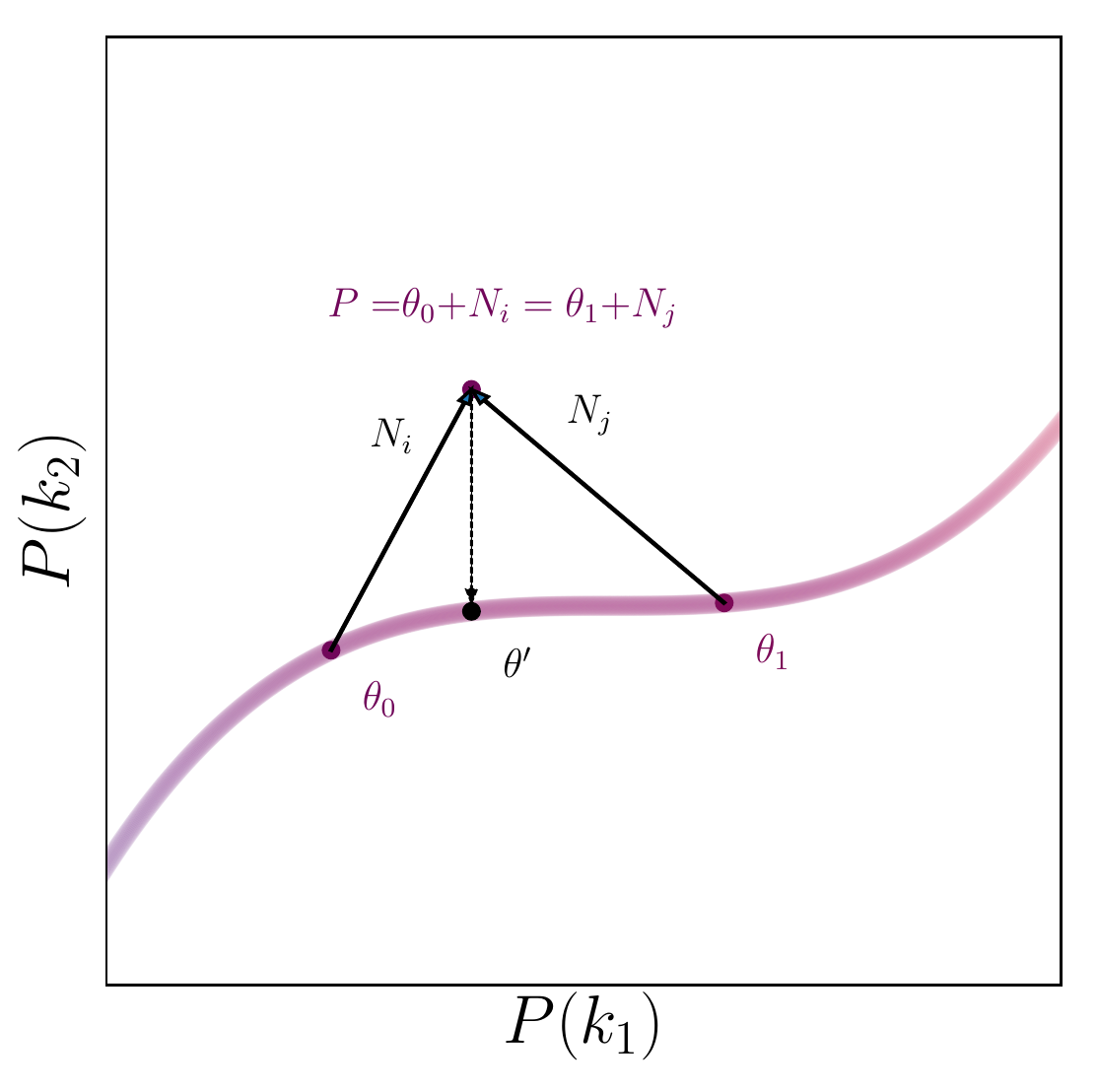}
  \includegraphics[width=.45\columnwidth]{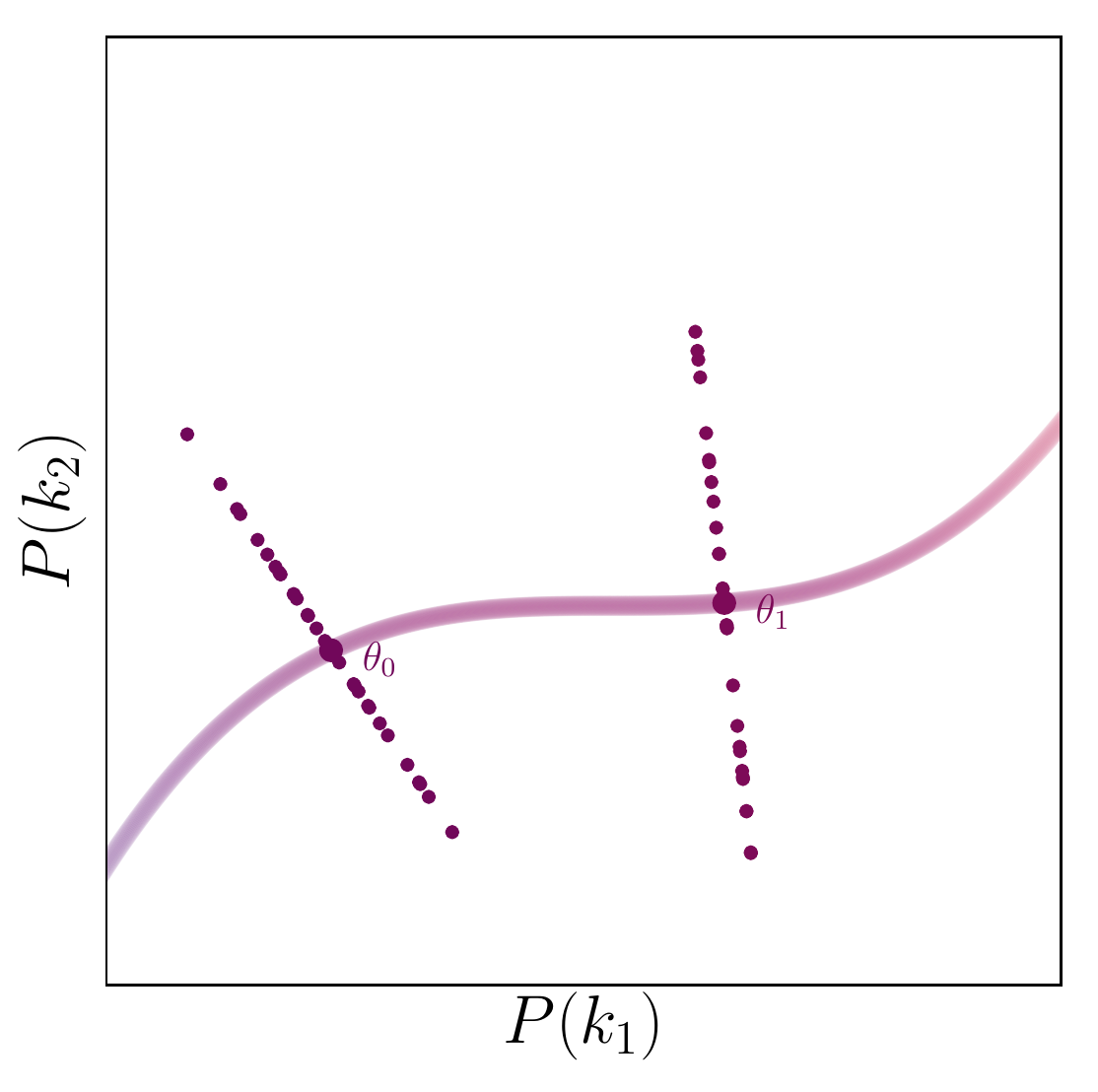}
\caption{A toy model with a single parameter $\theta$ predicting a 2-valued power spectrum. The left panel shows how, when adding noise, two different parameter values can result in the same noised signal, neither of which would be the highest-likelihood parameter value ($\theta^\prime$ in this case). The right panel shows how the ambiguity disappears when considering noise realizations that are perpendicular to the model-manifold. }
\label{Fig:PerpProblem}
\end{figure}

	A difficulty appears when using a noised signal. For a sufficiently dense learning set, the signals corresponding to two neighbouring cases, noised with two different realizations of the instrumental thermal noise, could lead to the same noised signal. We show this problem in the left panel of figure \ref{Fig:PerpProblem} for a toy model with a 2D signal, composed of the power spectrum values at two wavenumbers $k_{1}$ and $k_{2}$, and produced by a model with only one parameter $\theta$. The purple line represents the manifold of all the possible signals produced by the model, the model-manifold for short. We illustrate that the noised signal $P$ has been obtained in two different ways, one starting from the signal corresponding to parameter value $\theta_{0}$ with a noise $N_{i}$ and the other starting from the signal corresponding to parameter value $\theta_{1}$ with a noise $N_{j}$. To correctly evaluate the prediction ability of the supervised learning methods, it is therefore necessary to specify what is the correct parameter value corresponding to a noised signal. The most logical answer is to decide that the correct parameter is the parameter corresponding to the un-noised signal on the model-manifold closest to the considered noised signal. If a natural distance is chosen in the signal space (e.g $L_2$ norm), this minimal distance corresponds to adding the most likely thermal noise. This is in essence the maximum-likelihood value for the parameter.
	In the toy model of figure \ref{Fig:PerpProblem}, we note this most probable parameter $\theta'$. When we give to our methods the noised signal $P$, we thus expect them to predict the parameter $\theta'$.

	Finding this most probable parameter value for a given noised signal is actually the very purpose of methods that derive parameter constraints. We do not know how to do it at low cost when building our noised training and test sets, at least in the general case. We address this issue by generating a perpendicularized noised signal whose corresponding most probable un-noised signal on the model-manifold is exactly one of the cases in our un-noised training or test sets. This noised signal belongs to a 117 dimensional hyper-plane of the signal space which is orthogonal to the model-manifold at the most-probable-signal's position. Graphically, we show a finite number of these perpendicularized noised signals for the two signal corresponding to $\theta_{0}$ and $\theta_{1}$ in the right panel of figure \ref{Fig:PerpProblem}. Note that mathematically this procedure does not fully lift the ambiguity: two perpendicular hyperplanes from two neighbouring points of the model-manifold will intersect on the concave side of a curved manifold. Conversely an observed noised signal (on the concave side) will belong to a single perpendicular hyperplane only if its distance to the model-manifold is smaller than the radius of curvature of the manifold at the intersection point (speaking in terms of the two-dimensional case).
	
\subsubsection{Generating the perpendicular noise}\label{SubSec:PerpAlgo}

	The perpendicularized noise realizations that have to be generated are perpendicular to the model-manifold at the position of the signal to which they will be added. Let us first consider the case when the signals have been generated at the nodes of a grid in the parameter space, like for our learning set. In this scenario, we can use finite differences to estimate the tangent hyper-plane to the model-manifold, and thus the dual perpendicular space.  For a signal $\mathbf{P}_{\Zion^{i},\Rf^{j},\Tvir^{k}}$ corresponding to the parameters $(\Zion^{i};\Rf^{j};\Tvir^{k})$, where $i,j$ and $k$ denote the indexes on the grid, we apply an algorithm whose main steps are the following :
\begin{enumerate}
	\item Compute the vectors
	\begin{eqnarray}
	 \mathbf{V}_{\Zion ,i,j,k}&=&\mathbf{P}_{\Zion^{i+1},\Rf^{j},\Tvir^{k}}-\mathbf{P}_{\Zion^{i-1},\Rf^{j},\Tvir^{k}}\\ 
	 \mathbf{V}_{\Rf ,i,j,k}&=&\mathbf{P}_{\Zion^{i},\Rf^{j+1},\Tvir^{k}}-\mathbf{P}_{\Zion^{i},\Rf^{j-1},\Tvir^{k}}\\
	 \mathbf{V}_{\Tvir ,i,j,k}&=&\mathbf{P}_{\Zion^{i},\Rf^{j},\Tvir^{k+1}}-\mathbf{P}_{\Zion^{i},\Rf^{j},\Tvir^{k-1}}
	 \label{Eq:VectLocalBasis}
	\end{eqnarray} 
	  that form a local basis generating the hyperplane tangent to the model-manifold at signal $\mathbf{P}_{\Zion^{i},\Rf^{j},\Tvir^{k}}$.
	\item Orthonormalize the previous basis to obtain an orthonormal basis whose elements will be referred as $\mathbf{e}_{1,i,j,k}$, $\mathbf{e}_{2,i,j,k}$ and $\mathbf{e}_{3,i,j,k}$
	\item Generate a thermal noise $\mathbf{N}$ and compute
	\begin{equation}
		\mathbf{N}_{\perp}=\mathbf{N}-\mathbf{N}.\mathbf{e}_{1,i,j,k}-\mathbf{N}.\mathbf{e}_{2,i,j,k}-\mathbf{N}.\mathbf{e}_{3,i,j,k}
	\label{Eq:PerpendicularNoise}
	\end{equation}
	where $.$ denotes the standard euclidean scalar product in signal space.
\end{enumerate}

	Any such $\mathbf{N}_{\perp}$ is locally perpendicular to the model-manifold. Therefore, by construction, $(\Zion^{i};\Rf^{j};\Tvir^{k})$ are the parameters corresponding to the signal of the model-manifold closest to the noised signal $\mathbf{P}_{\Zion^{i},\Rf^{j},\Tvir^{k}}+\mathbf{N}_{\perp}$. 
	
	The second case to consider is the case when the signal to noise does not correspond to nodes of a grid in parameter space, like in our test set. In this case we determine the tangent hyper-plane using a weighted average of the hyper-planes for the neighbouring
	nodes of the grid. The details of the procedure are described in the appendix.

	It is worth noting that, with our method, only the local basis at signals which are not at the edges of our domain can be computed, therefore limiting the number of data in our learning set to $25920$ and in our test set to $2580$.

\subsubsection{Data pre-processing}

    As already mentioned in sec. \ref{SubSec:PerpTh}, considering the component of the noise perpendicular to the model-manifold does not completely lift the ambiguity in finding the signal on the model-manifold closest to a given observed signal, because of the manifold curvature. If the distance to the manifold is larger than the manifold inverse curvature (radius of curvature in 2D) there may be several points (or even a continuity of points) on the manifold whose perpendicular hyperplane goes trough the noised signal, only one of them being the closest. Thus our method inherently has difficulty with observed signals far away from the manifold. If we are not careful, including a typical radio-interferometer thermal noise will automatically generate noised signals far from the manifold. Indeed, the model can easily (and does) generate
    the power-spectrum at large wavenumbers where the thermal noise is large. In the case of a typical SKA layout the thermal noise on the power spectrum typically increases as $k^3$ \citep[see e.g.][]{Koopmans15}. Then, including large wavenumbers without caution will break our approach. Indeed, our orthonormalized basis does not match the basis consisting of the $120$ Dirac functions centered on the $(k,z)$ values where our power spectrum is evaluated. On the Dirac functions basis, the large thermal noise components are localized on a few vectors of the basis. But when projected on the orthonormalized basis it contaminates all components. This problem would be alleviated if the noise already had components of similar amplitude on the Dirac functions basis.
    
    We arrive at the same conclusion by considering a general problem for supervised learning. If the fluctuations of a component of the signal are dominated by noise and not by variation due to changing values of the model parameter, this component will be of little help in constraining the parameters. The more relevant quantity to consider is of course the signal-to-noise ratio. This is equivalent to the traditional "inverse variance weighting" used in radio-astronomy imaging. With this operation, the contribution of the noise will be similar for all components of the preprocessed signal. Using this preprocessing step and feeding the result to the supervised learning methods, we give ourselves a better chance that the noised signal will remain close to the model-manifold. We indeed verified that this pre-processing step generally improves the accuracy of the predictions.

\subsection{Kernel regression}
Let us now describe advanced versions of linear regressions that we will use as supervised learning methods.
\subsubsection{Linear regression}

	When addressing an interpolation problem, one obvious, yet useful method that exists is the simple linear regression, which consists of performing the following minimization:

\begin{equation}
		\min\limits_{\alpha, \beta_{j}} \left[\sum_{i=1}^{N_{\text{S}}} \left(y_{i}- (\alpha + \sum_{j=1}^{N_{\text{D}}} \beta_{j}x_{i,j})\right)^{2}\right]
\label{Eq:Reg_LstSq}
\end{equation}
where $N_{S}$ is the number of cases in the learning set, $N_{D}$ is the dimension of the input and $y_{i}$ is the output. This method results in a global linear approximation of the interpolating function which is then used to predict the outputs of the data of the test set starting from their inputs. It is also equivalent to approximating the model-manifold with a single hyper-plane.

It is therefore logical to think that, for each data of the test set, a local linear approximation of the interpolated function around the considered input will give better results. This is equivalent to locally approximate the model-manifold with an hyperplane, that varies depending on the location.

\subsubsection{Kernel smoothing}

	Introducing locality in the regression leads the class of supervised learning methods called Kernel smoothing regression methods \citep{Hastie2001}. To achieve this, to each cases of the learning set participating in the regression we apply a weight:
\begin{equation}
	K_{\sigma}\left( x_{0},x_{i} \right) = \frac{1}{\sqrt{2\pi}}e^{-\frac{D(x_{0},x_{i})^{2}}{2\sigma^2}}
\label{Eq:Weight_Kernel}
\end{equation}
where $\sigma$ is a scale hyper-parameter describing the desired level of locality and$$D(x_{0},x_{i})=\sqrt{\sum_{j=0}^{N_{D}} (x_{0,j}-x_{i,j})^{2}}$$ is the distance in the signal space between the considered signal $x_{0}$ and the input signal $x_{i}$ of the learning set. The quantity to minimize is then

\begin{equation}
		\min\limits_{\alpha, \beta_{j}} \left[\sum_{i=1}^{N_{\text{S}}} K_{\sigma}\left( x_{0},x_{i} \right)\left(y_{i}- (\alpha + \sum_{j=1}^{N_{\text{D}}} \beta_{j}x_{i,j})\right)^{2}\right]
\label{Eq:Reg_Kern}
\end{equation}
The parameters $\alpha$ and $\beta_{j}$ now depend on $x_{0}$. 

\subsubsection{Global ridge kernel regression}

	If the kernel smoothing method considers the local information, it does nothing to deal with an eventual degeneracy of the problem. Imagine that two of the input values that constitute the signal are are perfectly correlated in the un-noised case when varying one of the model parameters. Then the prediction by the regression will be sensitive only to the average of the two $\beta_i$ coefficients corresponding to these two correlated values. The two $\beta_i$ could take very large values as long as the mean is correct. But then, when noise is added, which is uncorrelated at the two signal value, these two large $\beta_i$ value will induce a large variance in the predicted parameter values. What was here described in the case where noise is added can also occur when moving from learning set to test set as the two sets can exhibit different levels of correlation between two signal values.
	
	This problem can be alleviated by constraining the values of the coefficients, particularly relatively to one another\citep{Hanke1998,Calvetti2000,Hastie2001}: this is what the ridge regression is about.
	
    To implement this constraint we add a penalty term in our minimization which becomes
	
\begin{equation}
		\min\limits_{\alpha, \beta_{j}} \left[\sum_{i=1}^{N_{\text{S}}} K_{\sigma}\left( x_{0},x_{i} \right)\left(y_{i}- (\alpha + \sum_{j=1}^{N_{\text{D}}} \beta_{j}x_{i,j})\right)^{2}+\lambda\sum_{j=1}^{N_{\text{D}}} \beta_{j}^{2}\right]
\label{Eq:Reg_RidgeKern}
\end{equation}
where $\lambda$ is an adjustable hyper-parameter. This ridge regression basically shrinks the values of the coefficients by imposing a penalty on their size. The whole coefficient shrinkage process is comparable to the weight decay process used in neural networks and we can assimilate the hyper-parameter $\lambda$ to a decay rate.

\subsubsection{Local ridge kernel regression}\label{SubSec:LocalRKR}

	One last step toward designing the most efficient regression is to consider the optimization of the two hyper-parameters $\sigma$ and $\lambda$. It is likely that a global optimization of the hyper-parameters values on the overall domain of the signal space will result in a selection of mean values which enable most of the space to be correctly predicted but might critically fail in some area of the domain. One simple improvement is therefore to determine the best hyper-parameters values for each points of the test set, thus minimizing the quantity of Equation \ref{Eq:Reg_RidgeKern} for $\alpha$, $\beta_{j}$, $\sigma$ and $\lambda$.

\begin{figure}
\begin{center}
	\includegraphics[width=\columnwidth]{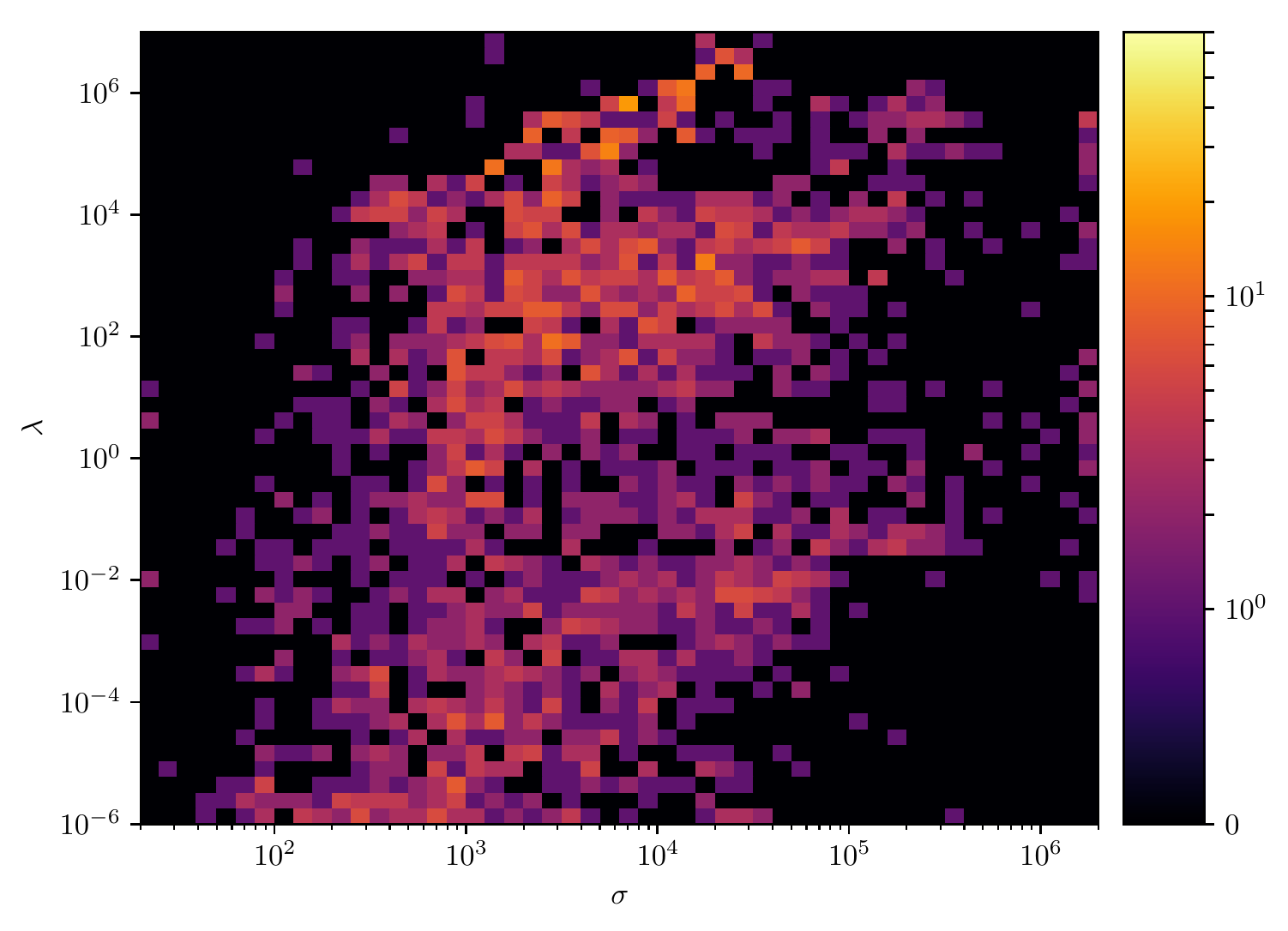}
	\caption{Histogram of the number of cases in the test set composed of noised signal, that find  a particular $(\sigma,\lambda$) duplet of hyper-parameters as being optimal for the prediction by the local ridge kernel regression method.}
	\label{Fig:HyperParam_Dispersion}
\end{center}
\end{figure}

    We implemented the local optimization by doing a simple grid search where we allowed our hyper-parameters to vary in a vast range of value:
    \begin{itemize}
	 \item[$-$]$\sigma\in [2\times 10^{1}, 2\times 10^{6}]$\\
	 \item[$-$]$\lambda \in [1\times 10^{-6}, 1\times 10^{7}]$\\
    \end{itemize}
    This wide range appeared to be necessary as the hyper-parameters indeed took vastly different values depending on the location in the signal space. To illustrate this assertion, we present in figure \ref{Fig:HyperParam_Dispersion}  an histogram of the optimized hyper-parameters values computed from all the cases in the test set composed of noised signals.
	
	However, whether locally or globally, optimizing the value of the hyper-parameters when predicting the outputs of the test set requires us to already know the true value of the outputs to compute an error function between the predictions and the true results. This will not be possible with an observed signal whose associated model parameters are unknown. Still, by using this knowledge in the case of the test set, we exhibit the theoretical maximum accuracy of this method. The optimization of the hyper-parameters in a real case (i.e. with an observed signal) is an open problem but one reasonable solution is to adopt for the same hyper-parameters as for the closest signal (in signal space) in the learning set. An optimization of the value of the hyper-parameters for each point of the learning set is therefore needed. Obviously, the accuracy of the predictions on the test set will be worse when using hyper-parameter values optimized for the closest signal in the learning set than when optimizing on the test set signal itself. This gap may be reduced in the future by improving the strategy to chose the optimal hyper-parameters of an observed signal.

\subsection{Artificial neural network}
\subsubsection{Network architecture}

\begin{figure}
\begin{center}
	\includegraphics[width=\columnwidth]{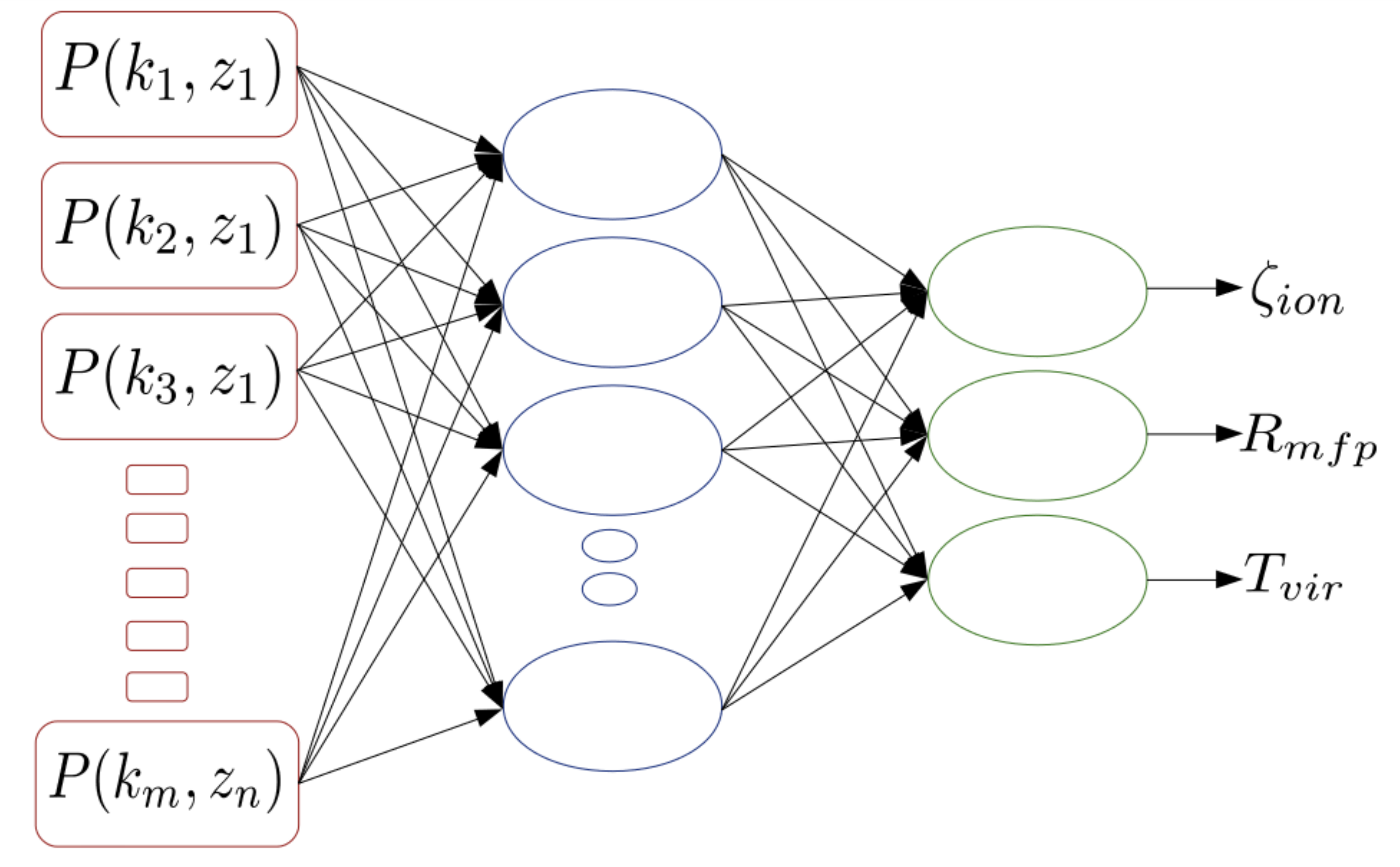}
	\caption{Schematized architecture of the neural network used in this study.}
	\label{Fig:NN1}
\end{center}
\end{figure}

	The principle of neural networks will not be discussed in depth as it has already been extensively described in various works \citep[for examples in this field, see][]{Shimabukuro2017,Jennings2018} but we remind the reader that a neural network is basically composed of a set of calculus units, called neurons, that return an output which is the value of a function, called activation function, acting on the weighted sum of the inputs to the neuron. These units can be linked together in numerous fashion defining the architecture of the network. In our case, we use the Keras framework\footnote{https://keras.io} relying on Tensorflow\footnote{https://www.tensorflow.org/} as a backend to implement a fully-connected  neural network with only one hidden layer of neurons, as shown in figure \ref{Fig:NN1}. Our hidden layer is composed of $80$ neurons and our output layer of $3$ neurons that each predicts the value of one of our three astrophysical parameters. With this simple architecture, a predicted parameter $y^{\text{pred}}$ can be explicitly written in term of the inputs to the network as :

\begin{equation}
	y^{\text{pred}}=f_{2}\left(\sum\limits_{i=1}^{80}W_{i}f_{1}\left(\sum\limits_{j=1}^{N_{D}}w_{i,j}x_{j}+b_{i}\right)+b\right)
\label{Eq:PredictedParameter}
\end{equation}
	where $N_{D}$ is the dimension of the input, $x_{j}$ the value of the $j$-th component of the input, $w_{i,j}$ the weight given to $x_{j}$ by the $i$-th neuron of the first layer, $b_{i}$ the bias added by the $i$-th neuron, $f_{1}$ the activation function of the first layer, $f_{2}$ the activation function of the second layer, $W_{i}$ the weight given by the neuron of the second layer that predicts $y^{\text{pred}}$  to the result of the $i$-th neuron of the first layer and $b$ the bias of the neuron of the second layer predicting the parameter $y^{\text{pred}}$. 

	We chose a fairly simple architecture for our network as it is fully-connected and contains only one hidden layer. However it has been mathematically proven \citep{Cybenko1989, Hornik1989} that neural networks with only one hidden layer can approximate with any accuracy any function if a sufficiently large number of neuron is used. The limit is rather in the size of the learning set describing the function to be interpolated. A network with many neurons but a small learning set will perform well on the learning set but weakly on the test set (overfitting phenomenon, or bias-variance trade-off).
	Evidences of the ability of this kind of network to handle the underlying denoising task have also been established \citep{Burger2012}.
	
\subsubsection{Network characteristics}
	The learning process of a neural network is done by a gradient descent algorithm, referred as optimization algorithm. During the learning process, the optimization algorithm, using a prediction and its quality estimated by an error function, adjusts the weights given by each neurons to each inputs. The newly adjusted weights are then used to make another prediction during the next step and the weights are again re-adjusted to minimize the error function. This scheme is repeated over thousands of step, called epochs of learning, until the error function stop decreasing. For reference, we present all the characteristics of our chosen learning process in Table \ref{Table:NN1_features}.

\begin{table*}
	\caption{The detailed characteristics of our neural network. $N_{TS}$ is the number of data in the test set, $N_{P}$ the number of predicted parameters, here 3, $y_{i,j}$ is the $j$-th output parameter for the $i$-th data in the test set, $t$ is the epoch of learning. $w^{t}_{k,l}$ is the weight given by the $l$-th neuron of a given layer to its $k$-th input at the epoch of learning $t$}
	\begin{tabular}{l c}
	\hline
	\hline
	\textbf{Characteristics Function} &  \\
	$\begin{array}{c}
	\text{Cost Function :}\\
	\text{Mean Squared Logarithmic Error}\\
	\end{array}$
	& $C=\frac{1}{N_{\text{TS}}}\sum\limits_{i=1}^{N_{\text{TS}}}C_{i}=\frac{1}{N_{\text{TS}}}\sum\limits_{i=1}^{N_{\text{TS}}}\frac{1}{N_{p}}\sum\limits_{j=1}^{N_{p}}\left[ \log_{10}\left(\frac{y_{i,j}^{\text{pred}}+1.}{y_{i,j}^{\text{true}}+1}\right)\right]^{2}$\\
	& \\
	\hline
	$\begin{array}{c}
	\text{Optimization Algorithm :}\\
	\text{RMSProp\citep{Murugan2017}}\\
	\end{array}$
	& 
	$\begin{array}{l}
	w_{k,l}^{t+1} = w_{k,l}^{t}-\frac{\eta}{\sqrt{E\left[(\nabla_{w_{k,l}}^{t}C)^{2}\right]_{t}+1\times 10^{-7}}}\nabla_{w_{k,l}}^{t}C \\
	\text{with }\\ E\left[(\nabla_{w_{k,l}}^{t}C)^{2}\right]_{t}=0.9 E\left[ (\nabla_{w_{k,l}}^{t-1}C)^{2}\right]_{t-1}+0.1(\nabla_{w_{k,l}}^{t}C)^{2}\\
	\end{array}$\\
	& \\
	\hline
	$\begin{array}{c}
	\text{Activation function of the hidden layer :}\\
	\text{ReLU}\\
	\end{array}$
	 &
	$f_{1}(x)=\left \{
   \begin{array}{r c l}
      x  & \text{if} & x>0 \\
      0   & \text{otherwise} & \\
   \end{array}
   \right .$\\
   \hline
  	$\begin{array}{c}
	\text{Activation function of the output layer  :}\\
	\text{Linear}\\
	\end{array}$
	& $f_{2}(x)=x$ \\
	\hline
	\hline 
	\textbf{Hyper-parameters} & \\
	Learning rate $\eta$ & $5\times 10^{-4}$\\
	\hline
	Batch size &  $128$\\
	\hline
	\end{tabular}
	\label{Table:NN1_features}
\end{table*}

\section{Results}

	Throughout this section, we will mainly quantify the quality of the predictions of our method through the root-mean square relative error, computed individually for each parameters and defined as :
\begin{equation}
	\chi_{y} =\sqrt{\frac{1}{N_{TS}}\sum\limits_{i=1}^{N_{TS}}\left(\frac{y_{i}^{\text{pred}}-y_{i}^{\text{true}}}{y_{i}^{\text{true}}}\right)^{2}}
\label{Eq:Khi}
\end{equation}
where $N_{TS}$ is the number of data in the test set and $y_{i}$ is the value of either $\Zion$,$\Rf$ or $\Tvir$ for the $i$-th case of this set. We first compare our different supervised learning method for the noise-free cosmological signal. Thus, we can compare our results to what has been done in \citet{Shimabukuro2017}. We then compare our methods on predicting the astrophysical parameters from a noised signal.

\subsection{Un-noised cosmological signal}
\subsubsection{Uncovering systematics}

\begin{figure}
\begin{center}
	\includegraphics[width=\columnwidth]{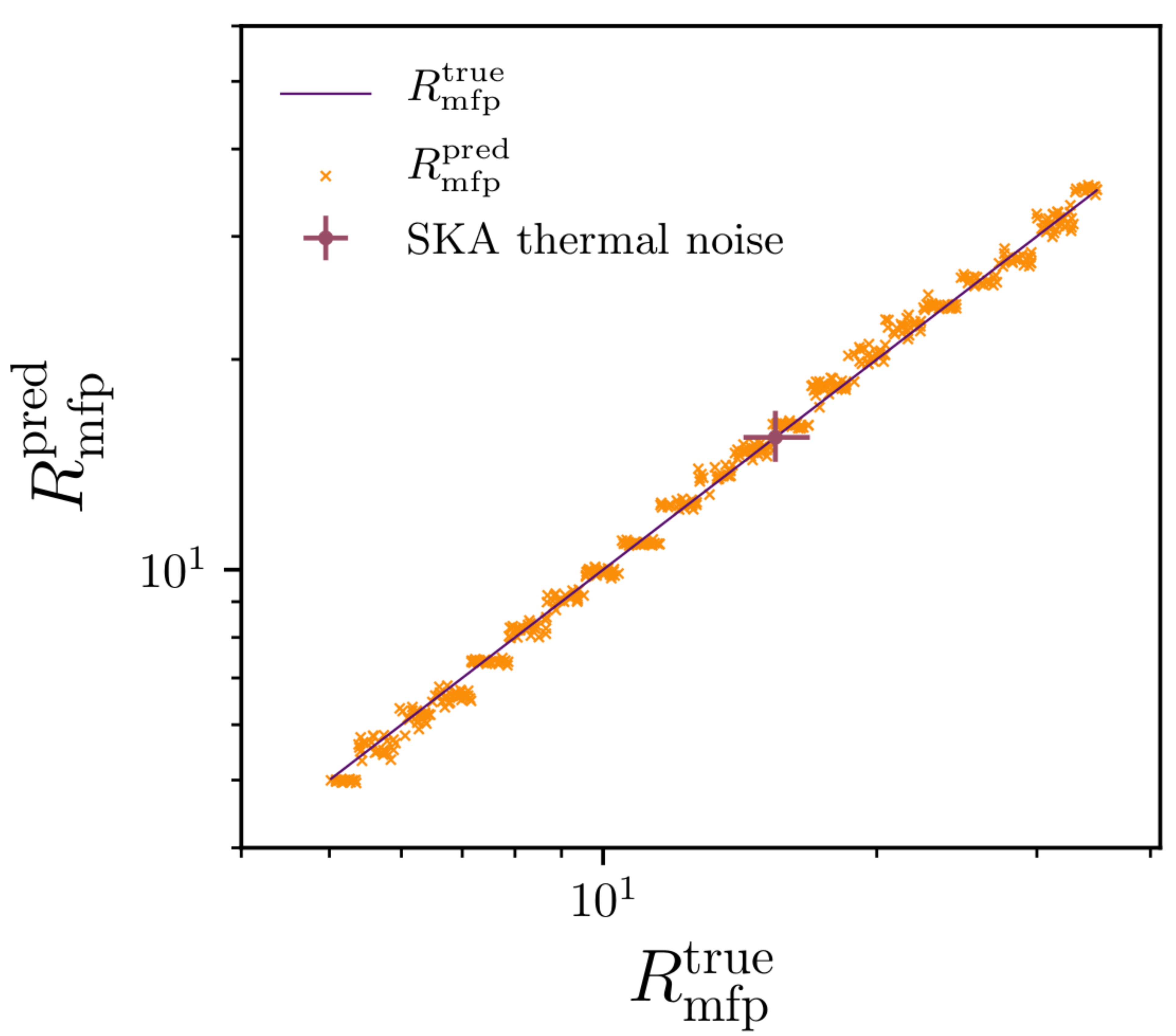}
	\caption{Predicted values of $\Rf$ as a function of the real ones for the theoretical perfect prediction (black line) and the actual predictions of our local ridge kernel regression method (yellow dots). For reference the typical amplitude of SKA thermal noise is also plotted (purple cross).}
	\label{Fig:Systematic_Rmfp}
\end{center}
\end{figure}

	During our investigation of the accuracy of the different the methods, we reached a systematic at some point. figure \ref{Fig:Systematic_Rmfp} depicts
	the predicted values of $\Rf$ as a function of the real ones, the black line being a perfect prediction and the yellow dot the actual predictions of our local ridge kernel regression method. It clearly shows that the predictions take mostly discrete values and are not simply exhibiting random deviations due around the value used to computed the signal. Looking into the 21cmFast source code, it appears that our method has correctly
	reconstructed a feature of the code which we did not take into account when labeling signals. In this version of 21cmFast, the radius of the regions to be tagged as
	ionized are investigated only in decrements of $\times 1.1$. Consequently only a variation of $\Rf$ by the same factor is guarantied to affect the
	results. More precisely, in our case, the influence of $\Rf$ is the same far all $\Rf$ in $\left[4.8903\times 1.1^{n},4.8903\times 1.1^{n+1}\right[$ for
	$n\in\mathbb{N}$, explaining the observed steps in figure \ref{Fig:Systematic_Rmfp}.

	For the rest of the study, we re-labeled our signals to correct this systematic by setting the "true" value of any $\Rf$ in a given interval to the geometrical mean value of the interval 
\begin{equation}
\begin{split}
	R_{\mathrm{mfp}}^{\mathrm{true}} & =\sqrt{4.8903\times 1.1^{n}\times 4.8903\times 1.1^{n+1}}\\
	& =4.8903\times 1.1^{n+\frac{1}{2}}
\end{split}
\end{equation}
for $n\in\mathbb{N}$. Not correcting for this systematic may have affected previous works attempting to constrain model parameters using 21cmFast. The resulting constraints may have been less tight than they could have been. Note that the last version of 21cmFast does not implement $\Rf$ in the same way and may not be equally sensitive to this systematic (although the discreet $\times 1.1$ factor remains).

\subsubsection{Performance comparison}

\begin{table*}
\caption{Root-mean square relative errors $\chi$ for a signal without noise, computed for each parameters, for  all the methods presented in Section \ref{Sec:Methods} where the regressions have been optimized using the test set. Results from \citet{Shimabukuro2017} are shown for comparison.}
	\begin{tabular}{c | l l l}
	\hline
	\hline
	 \textbf{Without noise} & $\chi_{\zeta_{\text{ion}}}$ & $\chi_{R_{\text{mfp}}}$ & $\chi_{log\left( T_{\text{vir}}\right)}$ \\
	\hline
	$\text{Shimabukuro 2017}$ & 27.1$\ODGcentieme$ & 22.8$\ODGcentieme$ & 2.7$\ODGcentieme$ \\
	\hline
	$\text{Linear Regression}$ & 1.82$\ODGcentieme$ & 8.00$\ODGcentieme$ & 0.29$\ODGcentieme$ \\
	$\text{Kernel Smoothing}$ & 1.19$\ODGcentieme$ & 6.13$\ODGcentieme$ & 0.28$\ODGcentieme$ \\
	$\text{Neural Network}$ & 1.37$\ODGcentieme$ & 2.53$\ODGcentieme$ & 0.15$\ODGcentieme$ \\
	$\text{Global Ridge Kernel Regression}$  & 0.68$\ODGcentieme$ & 1.76$\ODGcentieme$ & 0.09$\ODGcentieme$ \\
	$\text{Local Ridge Kernel Regression}$ & 0.39$\ODGcentieme$ & 0.19$\ODGcentieme$ & 0.04$\ODGcentieme$ \\
	\hline 
	\end{tabular}
	\label{Table:Khi_Unoised}
\end{table*}

\begin{figure*}
\begin{center}
	\includegraphics[width=2.\columnwidth]{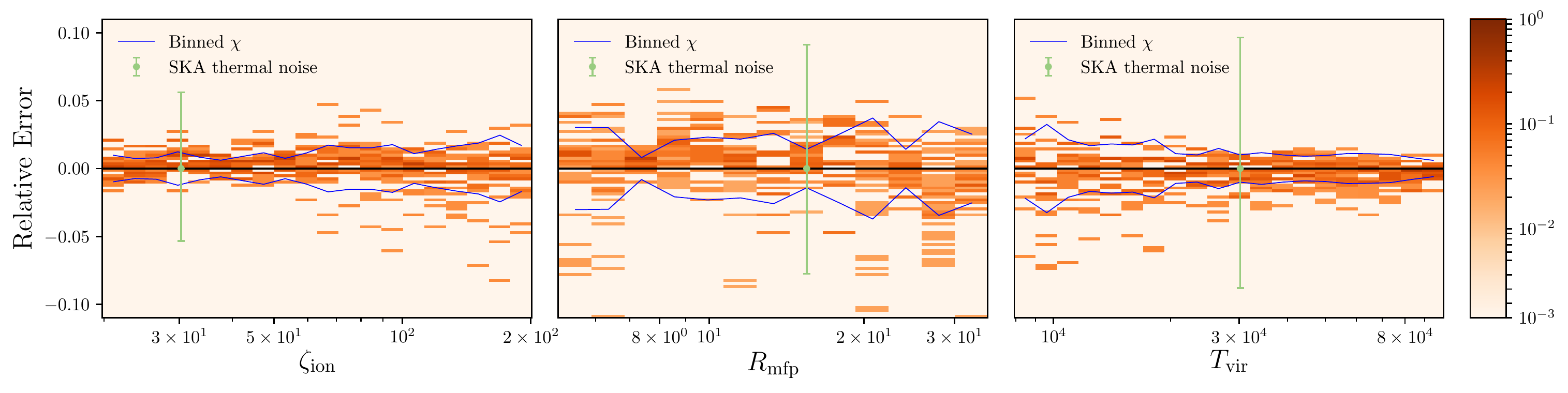}
	\includegraphics[width=2.\columnwidth]{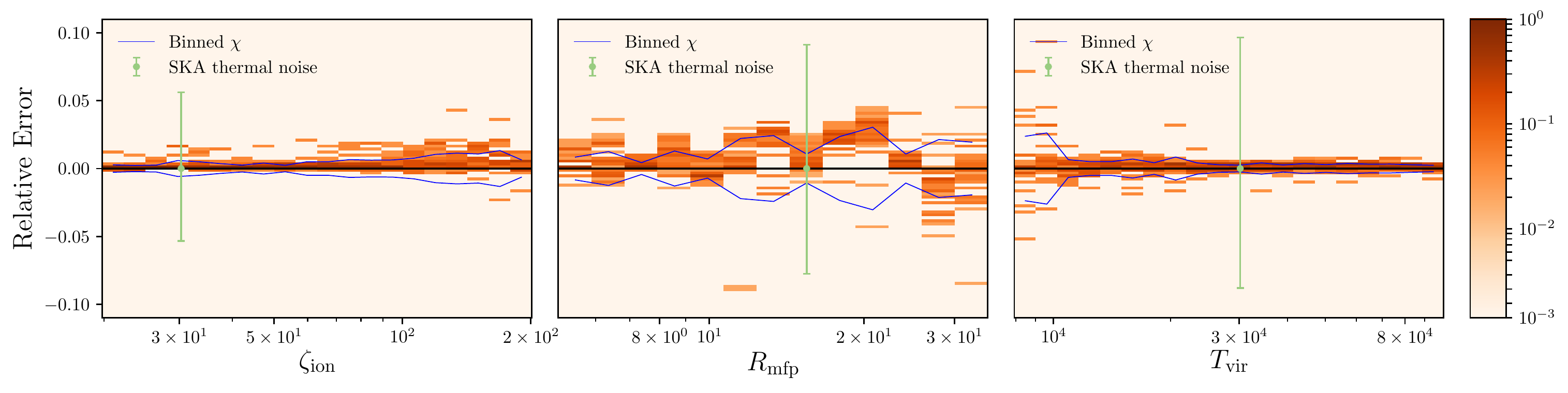}
	\includegraphics[width=2.\columnwidth]{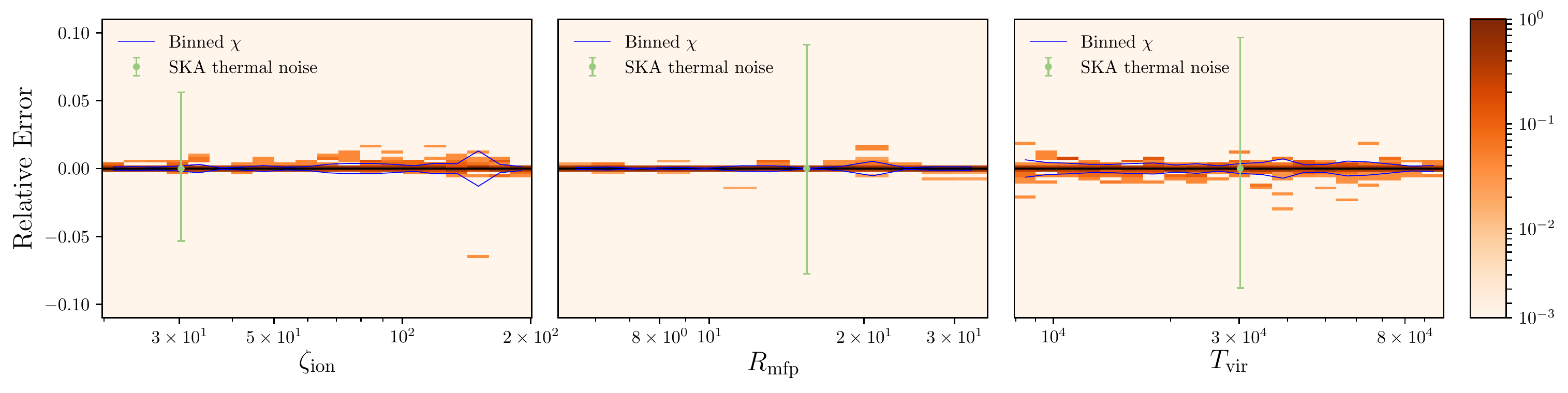}
\end{center}
\caption{Normalized distribution of the ratio $\dfrac{y_{\text{pred}}}{y_{\text{true}}}$ as a function of the true parameter $y_{\text{true}}$ for a signal without noises for our three astrophysical parameters $\Zion, \Rf$ and $\Tvir$.  The root-mean square error of the distribution is computed in different bins (blue line), and the 1$\sigma$ uncertainty from SKA thermal noise as estimated from Bayesian inference \citep{Greig2015a} is plotted for comparison (green point). All these are plotted for different supervised learning methods, from top to bottom : the neural network, the global ridge kernel regression and the local ridge kernel regression.}
\label{Fig:UnoiseResults}
\end{figure*}

	Table \ref{Table:Khi_Unoised} displays the root-mean square errors, computed individually for each parameters, for  all the methods presented in Section \ref{Sec:Methods} as well as results from \citet{Shimabukuro2017}. We find that all methods of this work is better by one order of magnitude than the results presented in \citet{Shimabukuro2017}. Beyond a more careful exploration and choice of the hyper-parameters of the learning process in the case of the neural network, the main reason for this improvement lies in the much larger learning set (justifying a network with more neurons) and the correction of the systematics on $\Rf$.
	Our best supervised learning method for predicting a cosmological signal appears to be the local ridge kernel regression which improved \citet{Shimabukuro2017} results by at least a factor $50$ for all three parameters, reaching a prediction rms relative error below $1\%$. However, remember that this result shows the theoretical maximum accuracy with perfectly optimized hyper-parameters.

	Figure \ref{Fig:UnoiseResults} shows the normalized distribution of the ratio $\dfrac{y_{\text{pred}}}{y_{\text{true}}}$ as a function of the true parameter $y_{\text{true}}$ for our three astrophysical parameters $\Zion, \Rf$ and $\Tvir$ as well as the the uncertainty induced by the SKA thermal noise as evaluated using Bayesian inference by \citet{Greig2015a} (green point) and the root-mean square error of the distribution (blue line). We present our three best supervised learning methods which are, from top to bottom : the neural network, the global ridge kernel regression and the local kernel ridge regression. For any of these methods, the root-mean square error of the predictions evaluated in different bins show only moderate fluctuations over the parameter range, and is inferior to a typical uncertainty induced by the SKA thermal noise. Moreover, for the local ridge kernel regression optimized on the test set the root-mean square error of the prediction is only a few percent of the uncertainty induced by the SKA thermal noise. While this needs to be confirmed in the case of the noised signal, it opens the door to using supervised learning as a method to determined the maximum likelihood parameters associated with an observed signal.

\subsection{Noised signal}

\begin{table*}
\caption{Root-mean square relative errors $\chi$ for a noised signal, computed for each parameters, for the methods presented in Section \ref{Sec:Methods} where the regressions have been optimized using the test set. The results have been shown for signals of the learning set noised with a perpendicularized noise. Results from \citet{Shimabukuro2017} are presented for comparison.}
	\begin{tabular}{c | l l l}
	\hline
	\hline
	 \textbf{With noise} & $\chi_{\zeta_{\text{ion}}}$ & $\chi_{R_{\text{mfp}}}$ & $\chi_{log\left( T_{\text{vir}}\right)}$ \\
	\hline
	$\text{Shimabukuro 2017}$ & 16.8$\ODGcentieme$ & 17.2$\ODGcentieme$ & 1.9$\ODGcentieme$\\
	\hline
	$\text{Neural Network}$ & 3.70$\ODGcentieme$ & 4.04$\ODGcentieme$ & 0.41$\ODGcentieme$ \\
	$\text{Global Ridge Kernel Regression}$  & 2.88$\ODGcentieme$ & 2.84$\ODGcentieme$ & 0.34$\ODGcentieme$\\
	$\text{Local Ridge Kernel Regression}$ & 1.10$\ODGcentieme$ & 0.60$\ODGcentieme$ & 0.16$\ODGcentieme$ \\
	\hline 
	\end{tabular}
	\label{Table:Khi_Noised}
\end{table*}

\begin{figure*}
\begin{center}
	\includegraphics[width=2.\columnwidth]{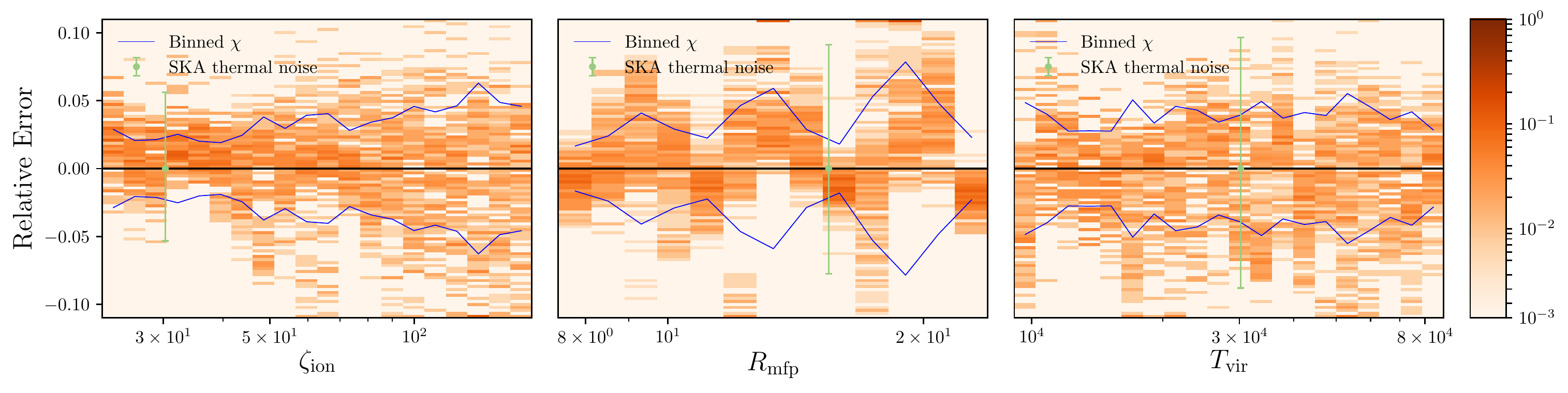}
	\includegraphics[width=2.\columnwidth]{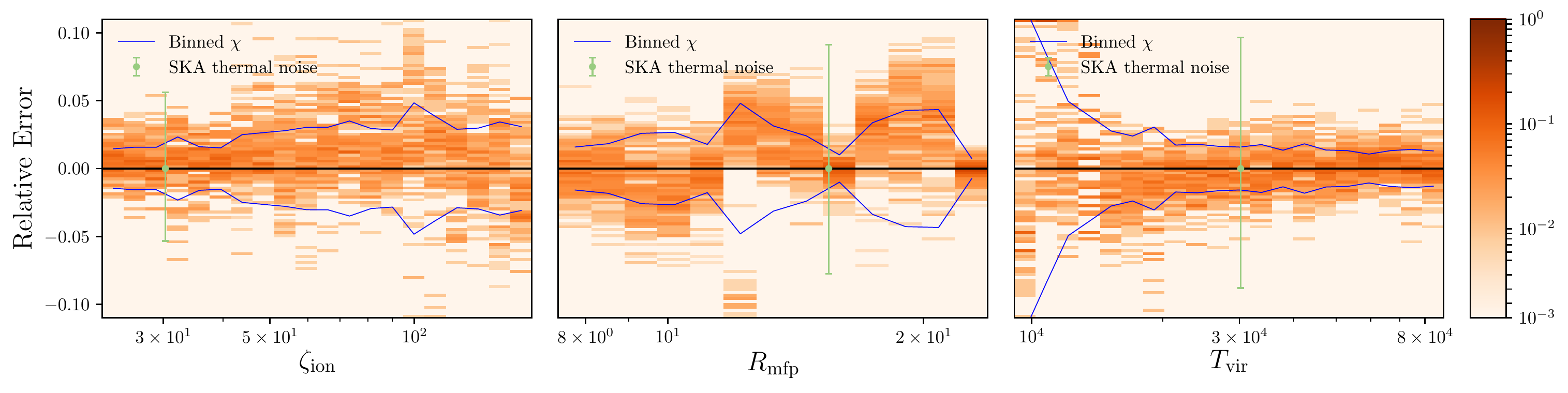}
	\includegraphics[width=2.\columnwidth]{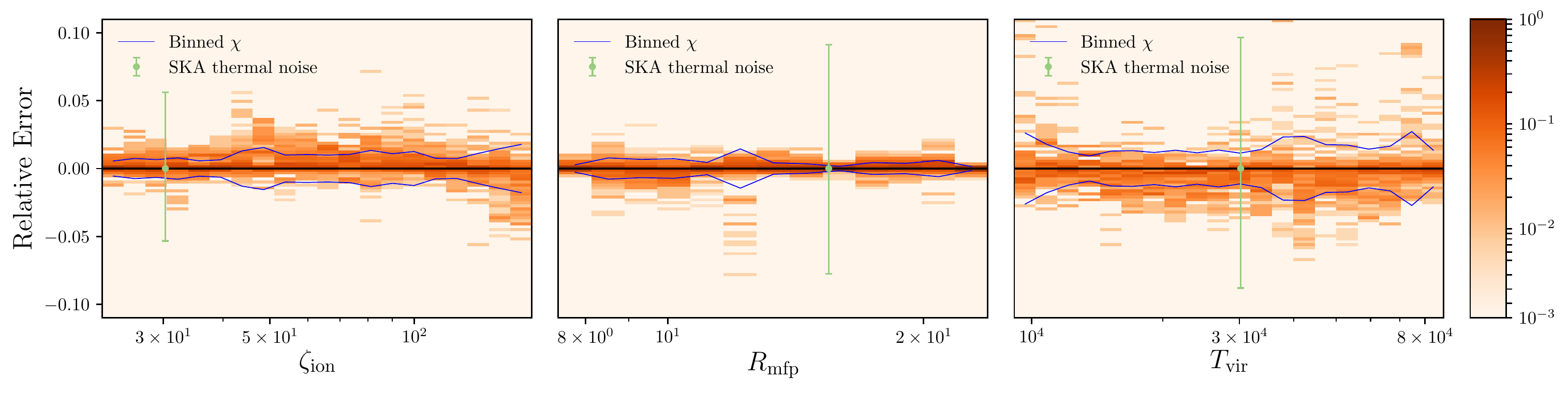}
\end{center}
\caption{
Normalized distribution of the ratio $\dfrac{y_{\text{pred}}}{y_{\text{true}}}$ as a function of the true parameter $y_{\text{true}}$ for a noised signal, using a learning set with perpendicularized noise,  for our three astrophysical parameters $\Zion, \Rf$ and $\Tvir$.  The root-mean square error of the distribution is computed in different bins (blue line), and the 1$\sigma$ uncertainty from SKA thermal noise as estimated from Bayesian inference \citep{Greig2015a} is plotted for comparison (green point). All these are plotted for different supervised learning methods, from top to bottom : the neural network, the global ridge kernel regression and the local ridge kernel regression.}
\label{Fig:NoiseResults}
\end{figure*}

\subsubsection{Perpendicularized noise for the learning set ?}

	As explained in Section \ref{SubSec:PerpTh}, the noise applied to the signals composing the test set has to be perpendicularized (with respect to the model-manifold) to enable a correct evaluation of the methods accuracy. This perpendicularization process also makes it necessary to pre-process the signals, weighting them by the inverse of the noise variance in each $k$ bin. In turn, this implies that the signals in the learning set should be weighted in the same way. However, the noise applied to the signals in the learning set can be either perpendicularized or not. If the case described in Section \ref{SubSec:PerpTh} and represented in figure \ref{Fig:PerpProblem} happens in the learning set, the learning process may be able to handle it by predicting an intermediate value between the two parameters. Using a perpendicularized noise would help the learning process by giving it directly the correct answer, but also generates a less uniform sampling of the signal space around the model-manifold. As it is difficult to decide in advance what is the stronger effect, we evaluated the quality of the predictions in the two cases: generic noise or perpendicularized noise applied to the learning set.
	

	We found that using a learning set with a perpendicularized noise leads to a slight improvement of the prediction accuracy. Even if not critical, we still decide to focus on the most accurate method and thus use a learning set with a perpendicularized noise. Let us now compare the performance of the different methods.

\subsubsection{Performance comparison}

\begin{table*}
\caption{Root-mean square relative errors $\chi$ for respectively : signals without noise (top) and noised signals (bottom). The $\chi$ have been computed individually for each parameters, for the global and local ridge kernel regression presented in Section \ref{Sec:Methods} and optimized using the learning set (LS) or the test set (TS). The signals of the learning set have been noised with a perpendicularized noise. Results for our neural network are shown for comparison.}
	\begin{tabular}{c | l l l}
	\hline
	\hline
	 \textbf{Without noise} & $\chi_{\zeta_{\text{ion}}}$ & $\chi_{R_{\text{mfp}}}$ & $\chi_{log\left( T_{\text{vir}}\right)}$ \\
	\hline
	$\text{Local Ridge Kernel Regression LS}$ & 2.69$\ODGcentieme$ & 3.79$\ODGcentieme$ & 0.48$\ODGcentieme$ \\
	$\text{Global Ridge Kernel Regression LS}$ & 1.36$\ODGcentieme$ & 3.78$\ODGcentieme$ & 0.30$\ODGcentieme$\\
	$\text{Neural Network}$ & 1.37$\ODGcentieme$ & 2.53$\ODGcentieme$ & 0.15$\ODGcentieme$ \\
	$\text{Global Ridge Kernel Regression TS}$  & 0.68$\ODGcentieme$ & 1.76$\ODGcentieme$ & 0.09$\ODGcentieme$ \\
	$\text{Local Ridge Kernel Regression TS}$ & 0.39$\ODGcentieme$ & 0.19$\ODGcentieme$ & 0.04$\ODGcentieme$ \\
	\hline
	\\
	\end{tabular}
	\begin{tabular}{c | l l l}
	\hline
	\hline
	 \textbf{With noise}& $\chi_{\zeta_{\text{ion}}}$ & $\chi_{R_{\text{mfp}}}$ & $\chi_{log\left( T_{\text{vir}}\right)}$ \\
	\hline
	$\text{Local Ridge Kernel Regression LS}$ & 10.3$\ODGcentieme$ & 14.1$\ODGcentieme$ & 4.23$\ODGcentieme$ \\
	$\text{Global Ridge Kernel Regression LS}$ & 2.89$\ODGcentieme$ & 2.83$\ODGcentieme$ & 0.34$\ODGcentieme$ \\
	$\text{Neural Network}$ & 3.70$\ODGcentieme$ & 4.04$\ODGcentieme$ & 0.41$\ODGcentieme$ \\
	$\text{Global Ridge Kernel Regression TS}$ & 2.88$\ODGcentieme$ & 2.84$\ODGcentieme$ & 0.34$\ODGcentieme$\\
	$\text{Local Ridge Kernel Regression TS}$ & 1.10$\ODGcentieme$ & 0.60$\ODGcentieme$ & 0.16$\ODGcentieme$ \\
	\hline 
	\end{tabular}
	\label{Table:Khi_OptLSTS}
\end{table*}

	From Table \ref{Table:Khi_Noised} we see that that all our methods are reconstructing the astrophysical parameters with an accuracy of the same order of magnitude from a noised signal as from a signal without noise, albeit worse by approximately a factor $3$. We show in figure \ref{Fig:NoiseResults} the normalized distribution of the ratio $\dfrac{y_{\text{pred}}}{y_{\text{true}}}$ as a function of the true parameter $y_{\text{true}}$ along with its root-mean square error (blue line) for our three astrophysical parameters $\Zion, \Rf$ and $\Tvir$. We show the results for our best methods which are respectively from top to bottom : the neural network, the global ridge kernel regression and the local ridge kernel regression. We also show the 1$\sigma$ uncertainty generated by SKA thermal noise as estimated with Bayesian inference \citep{Greig2015a} (green bars) for comparison. When considering the theoretical maximum accuracy of the local ridge kernel regression, which is when the hyper-parameters are optimized for the test set, we see that the prediction rms relative error is around $1\%$. Compared to SKA thermal noise, the reconstruction error is smaller by one order of magnitude. Thus, for a perfect optimization of the hyper-parameters, the local ridge kernel regression method enables a reconstruction of the highest likelihood astrophysical parameter with an error almost negligible compared to the size of the 1-sigma contour from Bayesian inference. 

\subsection{ Hyper-parameters optimization strategy}

	Previous results for the local regression method should be considered as the theoretical maximum accuracy. As explained in Section \ref{SubSec:LocalRKR}, we have optimized our hyper-parameters by using prior knowledge of the true value of the parameters, which is obviously not possible for an observed signal. Thus, we will now consider the case when the hyper-parameters have been optimized using only the information from the learning set, a method that can directly be applied to an observed signal. In this case, the hyper-parameters values assigned to a signal of the test set are the optimized values determined for the closest signal in the learning set. Let us note that the gap between the theoretical maximum accuracy and the accuracy of this optimization on the learning set may be narrowed in the future with better hyper-parameter optimization techniques.
	
\subsubsection{Un-noised cosmological signal}

	The top part of table \ref{Table:Khi_OptLSTS} displays the root-mean square relative errors, computed for each parameters, for a signal without noise using neural network and global and local ridge kernel regression optimized on the learning set (LS) or the test set (TS). For now, if we optimize the hyper-parameters with only the information from the learning set, the best method is the neural network. The Local Ridge Kernel regression accuracy worsen by a factor $\sim 10$, implying that it is sensitive to the value of the hyper-parameters, and that the optimal value are changing fast when moving away from the grid-point used in the regression.
	
\subsubsection{Noised signal}

	We also present in the bottom table of Table \ref{Table:Khi_OptLSTS} the root-mean square errors $\chi$, computed for each parameters, for a noised signal using neural network and global and local ridge kernel regression optimized on either the learning set (LS) or the test set (TS). The results are shown for signals of the test set noised with a perpendicularized noise. When not optimizing on the test set, the global ridge kernel regression is our most accurate way to reconstruct the astrophysical parameters with a prediction accuracy of a few percent or roughly half of SKA thermal noise. This method is barely affected by not using information from the test set, which is understandable since the optimization of the hyper-parameter is global. This simply states that the set test and learning set have comparable properties in this respect. We observe a decrease of the performence of the local ridge kernel regression when optimizing on the learning set similar to that in the case of un-noised signals.
	
	To summarize, the error on the prediction of the parameters caused by the supervised learning methods, although much improved, is not yet quite negligible compared to the SKA uncertainty when considering the method that could be directly applied to a the real signal. Algorithms to derive the optimal hyper-parameters to use on a real signal will have to be further improved. 
	
	
\section{Conclusion}\label{sec_ccl}

	In this work, we explored new supervised learning methods to constrain the underlying astrophysical parameters of the EoR. For this, we chose to base our reconstruction of the parameters on the power spectrum of the intergalactic 21-cm signal, measured at $12$ wavenumbers and each integer redshift from $z=5$ to $z=15$. We used 21cmFast to commpute the power spectra, varying three different parameters. We chose to vary $\Zion$ which accounts for the ionizing efficiency of high-z galaxies, $\Rf$ which is the mean free path of ionizing photons within the ionized regions and $\Tvir$ which expressed the minimum virial temperature for halos to be allowed to form stars.
	
	 We used a learning set of $2400$ signals produced by \citet{Eames2018}. They are generated on a 20$\times$6$\times$20 grid in the parameter space $(\Zion ; \Rf ; \Tvir )$. A test set of $512$ signals whose parameters are randomly picked within the bounds of the former set was also generated. To be more realistic we also analyze the case where a SKA-type thermal noise is added to the signals. It leads us to our first main result :

\begin{itemize}
	 \item The signal in the test set which is used to evaluate the prediction accuracy cannot be modified with a generic noise. If this is done, the most likely parameters values associated with the noised signals are unknown: they are not,in the general case, those that were used to produce the un-noised signal. Thus the accuracy of the prediction cannot be computed. To circumvent this issue, we have to perpendicularize the noise in the signal space relatively to the model-manifold.
\end{itemize}
	 
	We mainly implemented two supervised learning methods for our comparison. We first improved the neural network method by using a better optimization of the learning algorithm and hyper-parameters, and by using a larger learning sample, but we kept the  architecture from \citet{Shimabukuro2017} which is a fully-connected network with one hidden layer. Secondly, we studied another class of supervised learning methods which are different kinds of linear regressions and whose most advanced version is a ridge kernel regression with hyper-parameters optimized locally in the signal space. Comparing the prediction accuracy of those methods, we get the following results:

\begin{itemize}
	 \item For a 21-cm signal with no added noise, considering only methods which does not use information on the true value of the parameters to be predict to optimize its learning process, the best methods is the neural network. We predict the parameters with an error of a few percent, which is an order of magnitude better than in \citet{Shimabukuro2017}. On the other hand, if we focus on the theoretical maximum accuracy, the best method is the local ridge kernel regression whose hyper-parameters are optimized directly using information from the test set. This information would of course not be available in the case of an observed signal.
	 We find that, when the hyper-parameters are perfectly optimized, this methods leads to a prediction rms relative error below $1\%$, for a result $50$ times better than in \citet{Shimabukuro2017}.
	 \item When considering 21-cm signal with an added SKA thermal noise, the most accurate operational method is the ridge kernel regression globally optimized on the learning set with a prediction rms relative error of a few percent which is approximately half the amplitude of SKA thermal noise such as predicted in \citet{Greig2015a}. Again, from all methods the one with the theoretical maximum accuracy is the ridge kernel regression locally optimized which reconstruct the astrophysical parameters with an accuracy of the order of $1\%$ which is $10$ times lower than the predicted SKA noise amplitude, meaning that, once optimized to its maximum, this methods will recover the maximum likelihood astrophysical parameters with near negligible error due to the supervised learning method.
\end{itemize}

	As explained in Section \ref{SubSec:Evaluation_Issue}, our results are mitigated by the quality of our optimization of the hyper-parameters which we cannot prove to be a global optimization. Also, we optimize the performance of a neural network with only one hidden layer and do not explore the wide possibility of deep learning architecture with several hidden layer, which can very likely further improve the accuracy of the predictions. 

\section*{Acknowledgements}

This work was made thanks to the French ANR funded project ORAGE (ANR-14-CE33-0016). The simulations were performed on the GENCI national computing center at CCRT and CINES (DARI grants number 2014046667 and 2015047376). The authors also want to aknowledge F.Bolgar for its useful comments.

\bibliographystyle{mnras}
\bibliography{These-Supervised_Learning,myref}


\appendix
\section{Perpendicularized noise generation at generic parameter space location}
	To generate a noise perpendicular to the model-manifold at a location that is not on our initial sampling grid, we will still use the set defined on the grid as a way to obtain the local basis generating the hyper-plane tangent to the model-manifold. For a signal $\mathbf{P}_{\Zion^{x},\Rf^{y},\Tvir^{z}}$, our algorithm is:
\begin{enumerate}
	\item Identify the indexes $i$, $j$ and $k$ in the grid-generated set such that $\Zion^{i}\leq\Zion^{x}\leq\Zion^{i+1}$, $\Rf^{j}\leq\Rf^{y}\leq\Rf^{j+1}$ and $\Tvir^{k}\leq\Tvir^{z}\leq\Tvir^{k+1}$ which determine the eight signals from the grid-generated set that form the corners of the cell containing the considered signal $\mathbf{P}_{\Zion^{x},\Rf^{y},\Tvir^{z}}$.
	\item Using the algorithm for a grid-based set, compute the eight basis $\left( \mathbf{e}_{1,\alpha,\beta,\gamma};\mathbf{e}_{2,\alpha,\beta,\gamma};\mathbf{e}_{3,\alpha,\beta,\gamma}\right)$ with $\alpha=i$ or $i+1$, $\beta=j$ or $j+1$ and $\gamma=k$ or $k+1$, corresponding to these corner points.
	\item Compute the distances $D_{x,y,z}(i,j,k)$ between the considered signal and each of the eight previous points, based on the same definition of the scalar product in signal space.
	\item Compute a local basis at the considered signal $\mathbf{P}_{\Zion^{x},\Rf^{y},\Tvir^{z}}$ by making a weighted sum of the eight basis :
	\begin{eqnarray}
	 \mathbf{V}_{1,x,y,z}&=&\sum\limits_{\alpha =i}^{i+1}\sum\limits_{\beta =j}^{j+1}\sum\limits_{\gamma =k}^{k+1}W(\alpha,\beta,\gamma)\mathbf{e}_{1,\alpha,\beta,\gamma}\\ 
	 \mathbf{V}_{2,x,y,z}&=&\sum\limits_{\alpha =i}^{i+1}\sum\limits_{\beta =j}^{j+1}\sum\limits_{\gamma =k}^{k+1}W(\alpha,\beta,\gamma)\mathbf{e}_{2,\alpha,\beta,\gamma}\\ 
	 \mathbf{V}_{3,x,y,z}&=&\sum\limits_{\alpha =i}^{i+1}\sum\limits_{\beta =j}^{j+1}\sum\limits_{\gamma =k}^{k+1}W(\alpha,\beta,\gamma)\mathbf{e}_{3,\alpha,\beta,\gamma}
	 \label{Eq:VectLocalBasis}
	\end{eqnarray}
	where $$W(\alpha,\beta,\gamma)=\frac{\left[ D_{x,y,z}(
	\alpha,\beta,\gamma )\right]^{-1}}{\sum\limits_{\alpha '=i}^{i+1}\sum\limits_{\beta '=j}^{j+1}\sum\limits_{\gamma '=k}^{k+1}\left[ D_{x,y,z}(\alpha',\beta',\gamma' )\right]^{-1} }$$
	\item Orthonormalize the previous basis to obtain an orthonormalized basis whose elements will be referred as $\mathbf{e}_{1,x,y,z}$, $\mathbf{e}_{2,x,y,z}$ and $\mathbf{e}_{3,x,y,z}$
	\item Generate a Noise $\mathbf{N}$ and compute $\mathbf{N}_{\perp}$ using Equation \ref{Eq:PerpendicularNoise}.
\end{enumerate}

\end{document}